\documentclass[epj]{svjour}
\usepackage{graphics}
\usepackage{amsmath, amssymb}
\usepackage{tikz}
\usepackage{subcaption}

\begin{document}
%
\title{Chimera states in networks of logistic maps with hierarchical connectivities}
\author{Alexander zur Bonsen \and Iryna Omelchenko \and Anna Zakharova \and Eckehard Sch{\"o}ll
}                     
%
%
\institute{Institut f{\"u}r Theoretische Physik, Technische Universit{\"a}t Berlin, Hardenbergstr. 36, 10623 Berlin, Germany}
\date{Received: date / Revised version: date}
%
\abstract{
Chimera states are complex spatiotemporal patterns consisting of coexisting domains of coherence and incoherence. We study networks of nonlocally coupled logistic maps and analyze systematically how the dilution of the network links influences the appearance of chimera patterns. The network connectivities are constructed using an iterative Cantor algorithm to generate fractal (hierarchical) connectivities. Increasing the hierarchical level of iteration, we compare the resulting spatiotemporal patterns. We demonstrate that a high clustering coefficient and symmetry of the base pattern promotes chimera states, and asymmetric connectivities result in complex nested chimera patterns.  
\PACS{
      {PACS-key}{discribing text of that key}   \and
      {PACS-key}{discribing text of that key}
     } 
} 
\maketitle
\section{Introduction}
\label{sec:intro}

The dynamics of networks is a topic of great interest, offering applications to various fields such as technological, biological, neuronal, social systems, and  power grids \cite{SCH16,PAN15,WAT98}. A great variety of collective dynamics can be observed in networks, ranging from completely synchronized to spatially incoherent, desynchronized states. Chimera states are an intriguing partial synchronization pattern, first discovered in networks of identical phase oscillators with a symmetric coupling function \cite{KUR02a}. In spite of the symmetric configuration, symmetry-breaking collective behaviour can evolve spontaneously in such networks, namely a state in which certain nodes form spatially coherent, synchronized domains, while others move in a desynchronized, incoherent fashion.

Since their initial discovery in a system of phase oscillators, chimera states have sparked a large body of further work, e.g. \cite{MOT10,BOR10,OME10a,OME12a,MAR10,WOL11a,BOU14,SET14,FEN15,PAN15}. Chimera-like states were found in a variety of systems, among them time-discrete maps with chaotic and periodic dynamics \cite{OME11,SEM15a,BOG16,BOG16a,VAD16,SEM17,SHE17}, time-continuous chaotic systems \cite{OME12,SEM15a}, neural systems \cite{OME13,ROT14,OME15,SEM16,AND16,AND17}, Boolean networks \cite{ROS14a}, population dynamics \cite{HIZ15}, and quantum oscillator systems \cite{BAS15}. Besides these extensive numerical results, only one decade after their first observation, chimera states have been experimentally verified in several oscillator systems: optical \cite{HAG12}, chemical \cite{TIN12,NKO13}, mechanical \cite{MAR13,KAP14}, electronic and optoelectronic \cite{LAR13,LAR15} and electrochemical \cite{WIC13,WIC14,SCH14a}. 

Chimera states have also been found in networks with nonidentical oscillators \cite{LAI10} or with irregular topologies \cite{KO08,SHA10,LAI12,YAO13,ZHU14,SCH16b}, and in multilayer networks \cite{GHO16}. Even in globally coupled networks chimera states have been observed \cite{YEL14,BOE15,SCH14a}. A further type of coupling is inspired by recent findings in neuroscience, which revealed an inhomogeneous, fractal-like (hierarchical) topology of neurons in the human brain \cite{KAT09,KAT12,KAT12a,PRO12,EXP11a}. For such hierarchical, quasi-fractal connectivities chimera states have also been observed recently in various time-continuous systems \cite{OME15,HIZ15,ULO16,TSI16,SAW17}. The present work continues the systematic exploration of hierarchical connectivities in a network of time-discrete maps.

\section{Hierarchical connectivities}
\label{sec:hierC}

The motivation to study hierarchical connectivities in networks arises from recent findings in neuroscience, obtained mainly by the analysis of Diffusion Tensor Magnetic Resonance Imaging (DT-MRI) data. Rather than spreading homogeneously in three spatial dimensions, the axons of neurons in the human brain connect to other neurons in a hierarchical way \cite{KAT09,KAT12,KAT12a,PRO12,EXP11a}. Thus they only cover a fractal subset with typical fractal dimensions between 2.4 and 3 \cite{KAT12a}. This structure enables fast and optimal processing of information in the brain \cite{SCH16b}. 

The construction of the coupling scheme for the network is carried out by utilizing the fractal construction algorithm of a Cantor set as in ref. \cite{ULO16,TSI16}. An initiation string (base pattern) of zeros and ones, with a given length $b$ and the number of '1' given by $c_1$, e.g. 110111 $(b=6, c_1 = 5)$, defines the subdivision rule. In each iteration every 1 is replaced by the base pattern $b_{init}$ and every 0 with an equally long string of $b$ zeros. If repeated $n$ times, a hierarchical pattern of ones and zeros of length $b^n$ is created. The pattern defines the connectivity structure of the network, where each element represents either a link ('$1$') or a gap ('$0$'). The full hierarchical coupling pattern created hereby is denoted as $(b_{init})^n$. The fractal dimension $d_f = \ln c_1 / \ln b$ of the created hierarchical coupling pattern is an exact description for $n\rightarrow\infty$. As we use a finite number of steps repeating the Cantor algorithm, we call the created connectivity {\it hierarchical} or {\it quasi-fractal}. 

Using the resulting string as the first row of the adjacency matrix, and constructing a circulant adjacency matrix by applying this string to each element of the ring, a ring network of $N=b^n$ nodes with hierarchical connectivity is generated.
We slightly modify this procedure by adding an additional zero in the first instance of the sequence, which corresponds to the self-coupling. This procedure ensures the preservation of an initial symmetry of $b_{init}$ in the final link pattern, which is crucial for the observation of chimera states \cite{ULO16}, since asymmetric coupling leads to a drift of the chimera \cite{BIC15,OME16}. Thus a ring network of $N=b^n+1$ nodes is generated.

\begin{figure}[tb]
\includegraphics[width=\columnwidth]{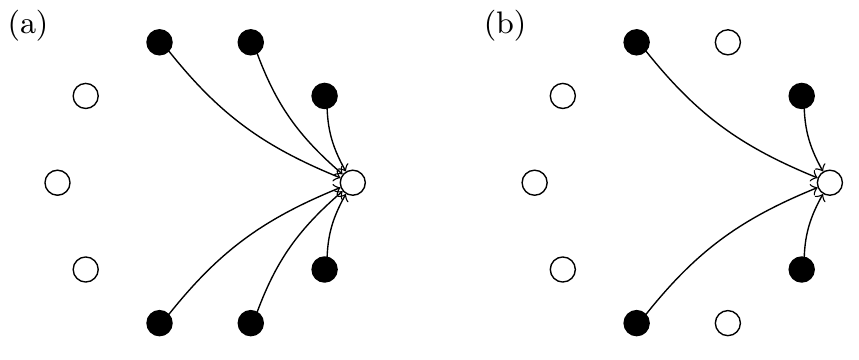}
\caption{Illustration of the network with $b_{init} = 101$, $n = 2$ for $N=10$, where only the links of a reference element are shown. (a) $m = 1$ results in non-local coupling, (b) $m=2$ results in hierarchical connectivity.}
\label{fig:network}
\end{figure}

To study the effect of increasing hierarchical connectivity in a system, we introduce the concept of hierarchical steps $m$. The hierarchical expansion of $b_{init}$ as described above, is interrupted after $m \leq n$ instead of $n$ steps. Afterwards, this pattern is expanded to the final size $N$ by replacing each element with $\frac{N-1}{b^m}$ copies of itself. 
For example, with $b_{init}=101$ the hierarchical expansion from $n=1$ to $n=2$ would result in 101000101. The homogenous expansion, used if $m=1$, results in 111000111, see Fig.~\ref{fig:network} for an illustration.

The algorithm still yields a topology of size $N = b^n+1$. But for $m < n$ the structure is less dilute and includes more links than the full hierarchical connectivity.
The number of links $c_{n,m} = \sum_{j=0}^{N-1} C_{0j}$, where $C_{kj}$ is the adjacency matrix of dimension $N\times N$, is determined by the number of links $c_1$ in $b_{init}$, the hierarchical level $n$ and step $m$. It is $c_{n,m} = c_1^mb^{n-m}$.

The algorithm allows for the construction of nonlocal topologies, too. They can be created by choosing a base pattern with the same number of links on each side, like 101 or 110011, and $m=1$ for any hierarchical level $n$, see Fig.~\ref{fig:network}. This permits to observe the impact of increasing hierarchical connectivity upon a nonlocal system. Since the latter networks have already been investigated \cite{OME11}, this offers an interesting reference point.

\section{The Model}
\label{sec:model}

The system under scrutiny is a ring of $N$ logistic maps:

\begin{alignat}{1}
  z_i^{t+1} = f(z_i^t) + \frac{\sigma}{c_{n,m}} \sum_{j=0}^{N-1} C_{ij} \left [f(z_j^t)-f(z_i^t) \right ].
  \label{eq:sys}
\end{alignat}
where $i=0,...,N-1$, all indices are modulo $N$, and the local dynamics of the individual nodes is governed by the logistic map  $f(z^t) = az^t(1-z^t)$ with bifurcation parameter $a$. Throughout this work we fix $a=3.8$, then each individual map operates in the chaotic regime.
The coupling strength is denoted by $\sigma$, and $c_{n,m}$ refers to the number of links in the topology. The coupling matrix $C_{ij}$ is constructed using the Cantor algorithm for fixed base pattern and hierarchical step, as described above.

\subsection{Clustering coefficient}
\label{sec:clustC}

Given various base patterns, a variety of different hierarchical connectivities can be constructed. These topologies may vary with respect to system size $N$, symmetry, link density $\rho(c_1, n, m) = c_{n,m}/(N-1)$, fractal dimension $d_f(b,c_1)$, and distribution of links and gaps. But differences in the latter are not captured by the current measures. Two hierarchical topologies with the same $b$, $n$, $m$ and $c_1$ but different $b_{init}$, e.g. 101 and 110, yield the same link density $\rho$ and fractal dimension $d_f$ but differ in their distribution of links and gaps, and their dynamics.

For this purpose the local clustering coefficient, as introduced by Watts and Strogatz \cite{WAT98}, is employed \cite{ULO16}. For a directed graph $G = (V,E)$ with set of nodes (or vertices) $V$ and set of edges $E$, it quantifies how strongly the neighbours of a node are interconnected. The closer a neighbourhood-graph is to a complete graph, a {\em clique}, the higher the clustering coefficient.

The neighbourhood $N_i$ of a node $i$ contains all nodes that the node connects to by an outgoing or incoming link: $N_i = \{v_j\!: e_{ij} \in E \lor e_{ji} \in E \}$. In directed graphs each direction of a link is considered separately. The maximum number of links in a neighbourhood of size $k_i = |N_i|$ is given by $k_i(k_i-1)$. The local clustering coefficient compares the existing number of links in the neighbourhood to the maximum possible number

\begin{alignat}{2}
C_i = \frac{|\{e_{jk}\!: v_j,v_k \in N_i, e_{jk} \in E\}|}{k_i(k_i-1)}.
\end{alignat}    

For circulant matrices the clustering coefficient $C_i$ is identical for all nodes and thus independent of $i$. Hence $C(b_{init}, n,m)$ denotes the clustering coefficient of a topology characterized by $b_{init}$, $n$ and $m$. It measures the compactness of the topology. The two extreme cases are global coupling with, e.g., $b_{init} = 111$, which yields $C(111,n,m) = 1$, and coupling to only one other nearest neighbor node, e.g. $b_{init} = 100$, which results in $C(100,n,m = n) = 0$.

In this work base patterns of length $b = 6$ with at least $c_1 = 3$ links are systematically examined. The latter restriction is made because it is assumed that chimera states are more likely to be found in networks with higher clustering coefficient. Consequently the network topology needs to contain a sufficient number of links (cf. Sect.~\ref{sec:clustC}). This choice offers a variety of 22 topologies, plus the global coupling case with $c_1 = 6$. This count excludes the mirror-images of non-symmetric strings. The mirror-image of a string has the same clustering coefficient and stability properties of the synchronized state. Hence only one representative of each non-symmetric pair is examined.

The hierarchy level is chosen as $n = 4$. Consequently the number of nodes equals $N = b^n +1 = 1297$ where an element is added to preserve initial symmetry of the base pattern. This size provides a reasonable compromise between numerical effort and validity of the results for real networks, e.g., neural networks. 

\begin{figure*}[tb]
    \centering   
    \includegraphics[width=\textwidth]{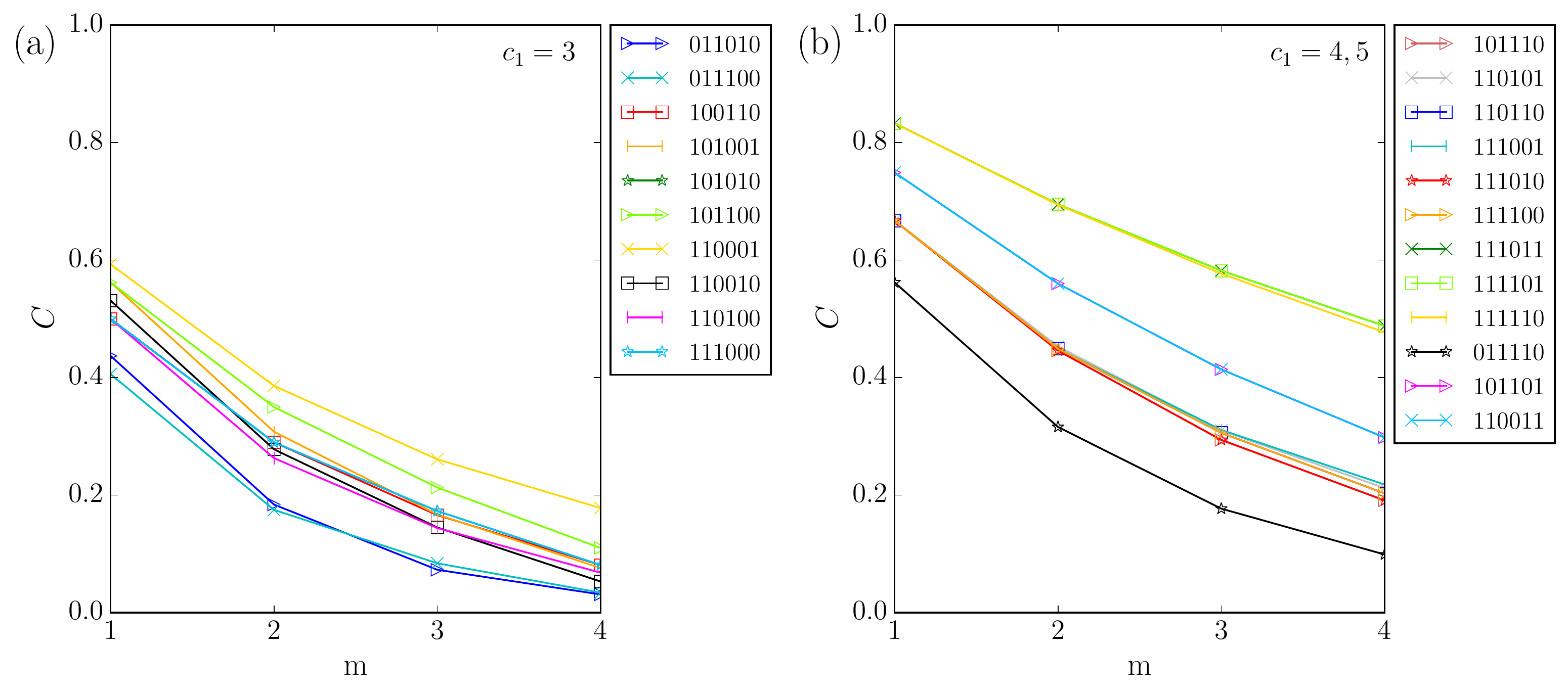}
    \caption{Clustering coefficients $C(b_{init},n,m)$ for all considered base patterns of length $b=6$ and hierarchical steps $n=4$, $N = 1297$. Left: $b_{init}$ with $c_1 = 3$. Right: $b_{init}$ with $c_1 = 4$ and $5$; last three strings in legend are symmetric}
    \label{fig:clust}
\end{figure*}

Figure \ref{fig:clust} presents the clustering coefficients of the considered base patterns. With increasing hierarchical stepsize $m$ the link density $\rho$ decreases. The remaining links become more irregular and possibly isolated. Hence it is expected that the clustering coefficient decreases with increasing $m$ as can be seen in Fig.~\ref{fig:clust}.

With one exception, the clustering coefficients increase with the number of links $c_1$, as is expected due to the higher link density. For strings with $c_1 = 5$ the coefficients are more or less the same. 
The base patterns with $c_1 = 4$ can be distinguished as mirror-symmetric and non-symmetric ones. The non-symmetric strings have similar clustering coefficients which slightly differ for $m>1$. Of the three symmetric strings, 011110 has lower and 110011, 101101 have higher and similar clustering coefficients. Symmetric strings result in symmetric coupling matrices, which has the consequence that all links in the network are bidirectional. Thus it should not be surprising that they have higher clustering coefficients. On the other hand, gaps at the end and beginning of the string result in lower clustering, as can also be found for $c_1 = 3$. This can be understood because the gaps increase the distance between a node and its neighbourhood. Consequently the overlap between the neighbourhoods of these neighbours is also smaller. This effect is big enough to cause the clustering coefficients of 011110 ($c_1 = 4$) to fall below the coefficients of strings with only $c_1 = 3$ links (110001, 101100).

On the other hand, the only other string with $c_1 = 3$ and links at each end, 110001, possesses the highest clustering coefficients of the strings with $c_1 = 3$. This stresses the decisive role of links and gaps at the ends of a string. But the internal structure, symmetry, and periodicity also have to be taken into account.

\section{Stability of the synchronized state}

\begin{figure*}[tb]
    \centering
    \includegraphics[width=\textwidth]{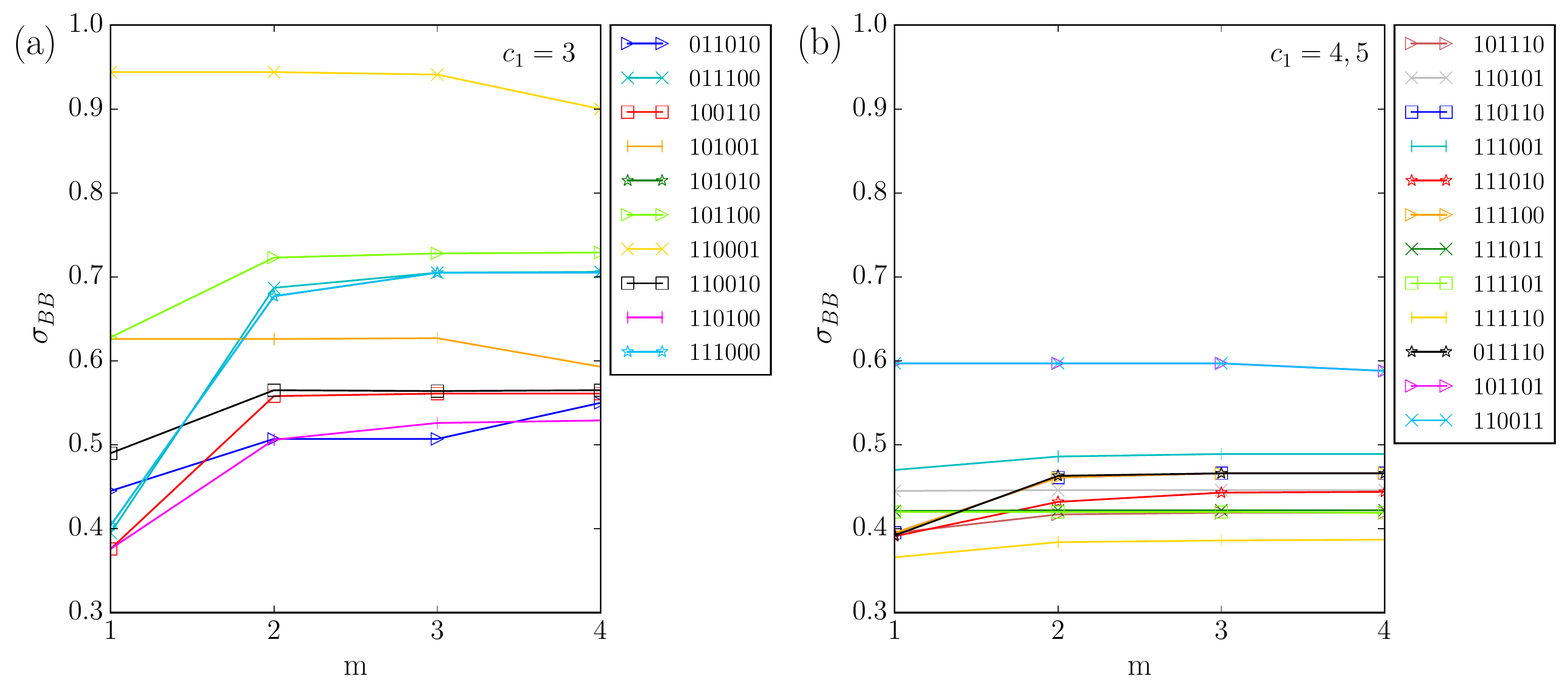}
    \caption{Minimum coupling strength for complete synchronization $\sigma_{BB}$ for all considered base patterns and hierarchical steps $n=4$, $N = 1297$. Left: $b_{init}$ with $c_1 = 3$. Right: $b_{init}$ with $c_1 = 4$ and $5$; last three strings in the legend are symmetric}
    \label{fig:sBB}
\end{figure*}

A regime of special interest is the parameter range of $\sigma$ in which the synchronized solution is stable. A linear stability analysis utilizing Lyapunov exponents, is used to obtain the blowout bifurcation $\sigma_{BB}$, i.e., the point where the synchronized state loses its transverse stability for coupling strength $\sigma < \sigma_{BB}$.

Figure \ref{fig:sBB} shows the results of the linear stability analysis for the 22 hierarchical base patterns investigated. The border of stability $\sigma_{BB}$ was computed for each hierarchical step $m = 1,2,3,4$. Above $\sigma_{BB}$ the synchronized state is stable. Hence, lower $\sigma_{BB}$ values correspond to a larger $\sigma$ interval in which the synchronized solution is stable. In the synchronized state the coupling term vanishes and the network evolves chaotically in time, because the logistic map is operating in the chaotic regime and mimics the dynamics of one uncoupled map.

Generally, increase of the hierarchical step, and consequently dilution of the links in the network, results in the decrease of the stability regions for the completely synchronized states. Larger values of coupling strength are required for the completely synchronized state to be stable.
In nonlocally coupled networks chimera states have usually been observed below the stability region of the synchronized states. Therefore, for fixed base pattern and hierarchical step, our interest will be focused on the regime of coupling strength $\sigma < \sigma_{BB}$.

Note that the increased stability of the synchronized state, which can be observed in some cases, is due to long-range links, introduced by the hierarchical connectivities, which extend through the whole ring and may even connect opposite sides (cf. \cite{TSI16}). Hence synchronized motion in a time-discrete system can be achieved by a smaller number of links in the network, if the nonlocal coupling of range $r$ is replaced by hierarchical connectivites with long-distance links.

\section{Chimera states}

The dynamics of the system depends on topology, bifurcation parameter $a$ of the logistic map, coupling strength $\sigma$, and initial conditions. To provide a systematic analysis, 88 topological alternatives are selected in this paper, resulting from 22 base patterns with four hierarchical steps. Mirror images of nonsymmetric patterns are not examined separateley, because numerical results show that the dynamics of mirror patterns is fairly similar.  Simulations are performed with initial conditions especially crafted to facilitate the observation of chimeras, as described in~\cite{HAG12}, as well as with random initial conditions. As in the case of nonlocally coupled networks with fixed coupling range, both choices result in chimera patterns for appropriate sets of system parameters, however, the specially prepared initial conditions allow for the observation of largest incoherent domains.

Figure \ref{fig:ChimOV} presents a selection of chimera patterns for various base patterns and various hierarchical steps. We show snapshots at fixed time (blue dots), and gray lines are used as a guide to the eye. In the upper row, chimera states with two incoherent domains are presented, in the middle row we demonstrate more complex chimera states with multiple incoherent domains, some of them have nested structure with small coherent domains inside the incoherent domains. The last row shows small- and large-amplitude incoherent domains, corresponding to two different types of chimera states: phase chimera and amplitude chimera, respectively~{\cite{BOG16,BOG16a}. While phase chimeras are associated with the spatially incoherent sequence of states on the upper and the lower branch of the snapshot profile (with a phase flip of $\pi$, i.e., half the period of the periodic dynamics of the coherent domains), amplitude chimeras denote incoherent small-amplitude modulations within one branch, similar to amplitude chimeras observed in time-continuous systems \cite{ZAK14,ZAK15b}.
In Fig.~\ref{fig:ChimOV}(i) the coexistence of phase and amplitude chimeras in the same system can be clearly seen.

These few examples demonstrate that the complexity of the patterns is a result of the interplay of the base pattern, the hierarchical step, the symmetry, and the coupling strength. To gain deeper insight into the mechanisms for the formation of chimera patterns, we will consider in more detail the dynamics of two exemplary base patterns: 110011 (symmetric) and 011100 (asymmetric).

\begin{figure*}[tb]
    \centering   
    \includegraphics[width=\textwidth]{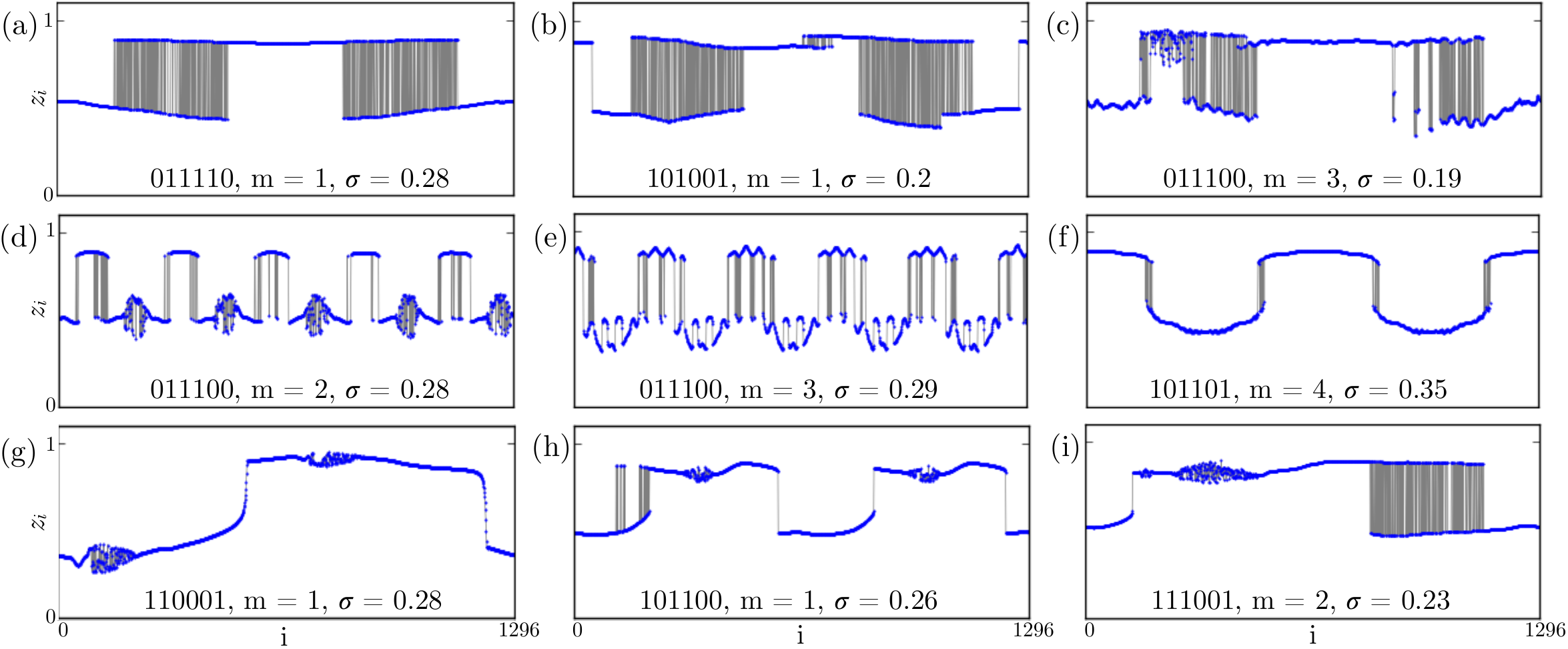}
    \caption{Selection of chimera patterns for various base patterns $b_{init}$ and hierarchical steps. Snapshots show last iteration $t=10000$. $b_{init}$, hierarchical step $m$ and coupling strength $\sigma$ are marked inside each panel; $n=4, a=3.8$.}
    \label{fig:ChimOV}
\end{figure*}

In nonlocally coupled networks of logistic maps~\cite{OME11}, for large coupling strength a coherent wave-like profile is stable and performs periodic dynamics in time. The wavenumber depends on the coupling range $r$, and smaller $r$ results in increasing wavenumber, i.e., more maxima and minima in these profiles.  When the coupling strength crosses the critical value $\sigma_{BB}$ and further decreases, the smooth profile is broken, and incoherent domains (phase chimeras) appear at the breaking points between the upper and lower branch. Consequently, a larger wavenumber of the coherent profile results in a larger number of incoherent chimera domains.
 Intuitively, one would expect a similar effect for higher hierarchical steps, since the number of links is decreasing;
however, long-range connections result in additional inhomogeneities giving birth to much more complex patterns.

Figure \ref{fig:110011stpprep} illustrates the dynamics of networks composed of the base pattern 110011 for $m=1,2,3,4$ with space-time plots for $20$ time steps. For all hierarchical steps, we observe a transition to chimera states with two incoherent domains. For higher hierarchical steps, the incoherent domains start to exhibit nested structures.

\begin{figure*}[tb]
    \includegraphics[width=\textwidth]{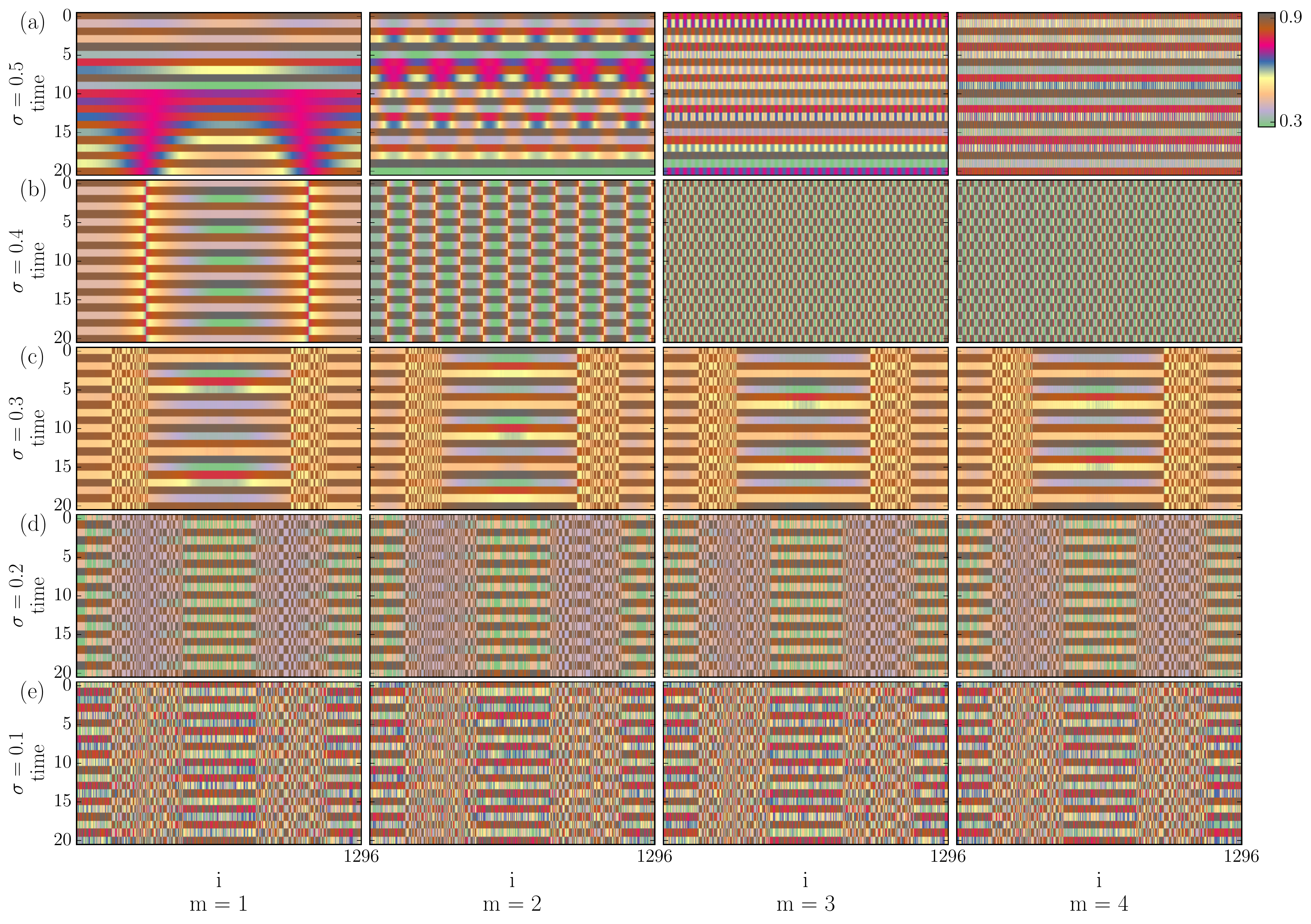}
    \caption{Spatio-temporal dynamics for $b_{init}= 110011$, $n = 4$, $N =1297$, $a = 3.8$ for various hierarchical steps $m$ and coupling strengths $\sigma$. Initial conditions are chosen by numerical continuation starting at $\sigma = 0.1$ with 
increment $\Delta\sigma = 0.01$ and iteration time $\Delta t = 10000$. The panels show space-time plots of the last 20 time steps. The rows (a)-(e) correspond to different values of $\sigma$, the columns correspond to different hierarchical stepsize $m$. 
}
    \label{fig:110011stpprep}
\end{figure*}



\begin{figure*}[tp]
\begin{minipage}[t]{0.48\textwidth}
\includegraphics[width=\textwidth]{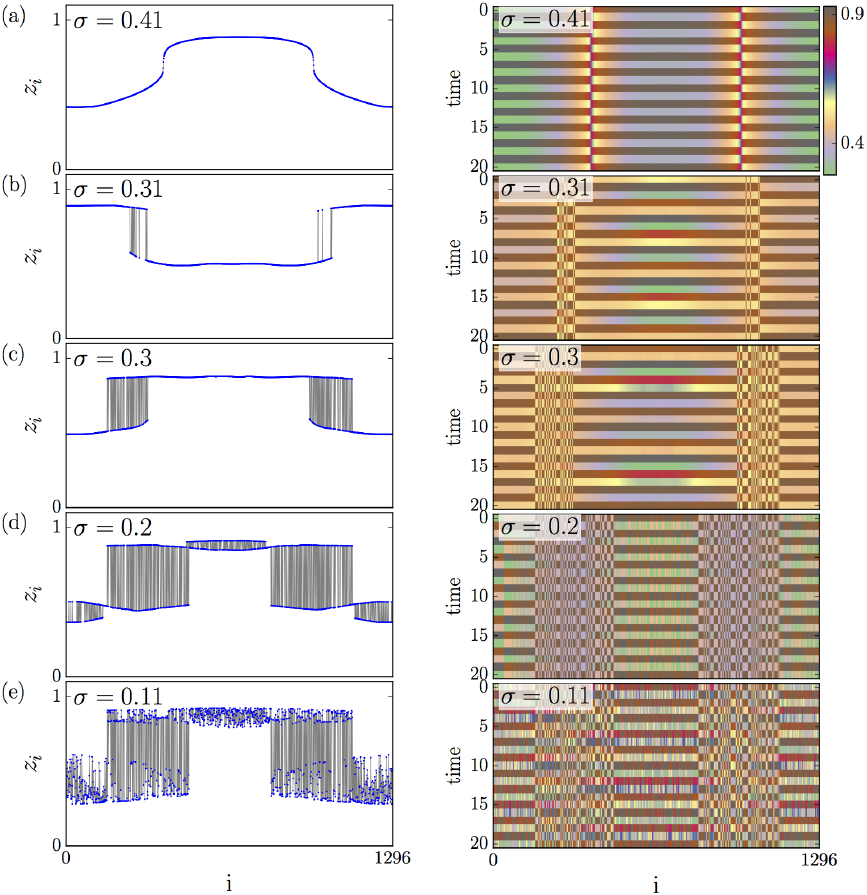}
\caption{Selected snapshots (left column) at t=10000 and space-time plots (right column) of last 20 iterations for base pattern $b_{init}= 110011$, $m=1$, $n = 4$. Further simulation details as in Fig.~\ref{fig:110011stpprep}.}
\label{fig:110011m14-I}
\end{minipage}
\begin{minipage}[t]{0.04\textwidth}
\end{minipage}
\begin{minipage}[t]{0.48\textwidth}
\includegraphics[width=\textwidth]{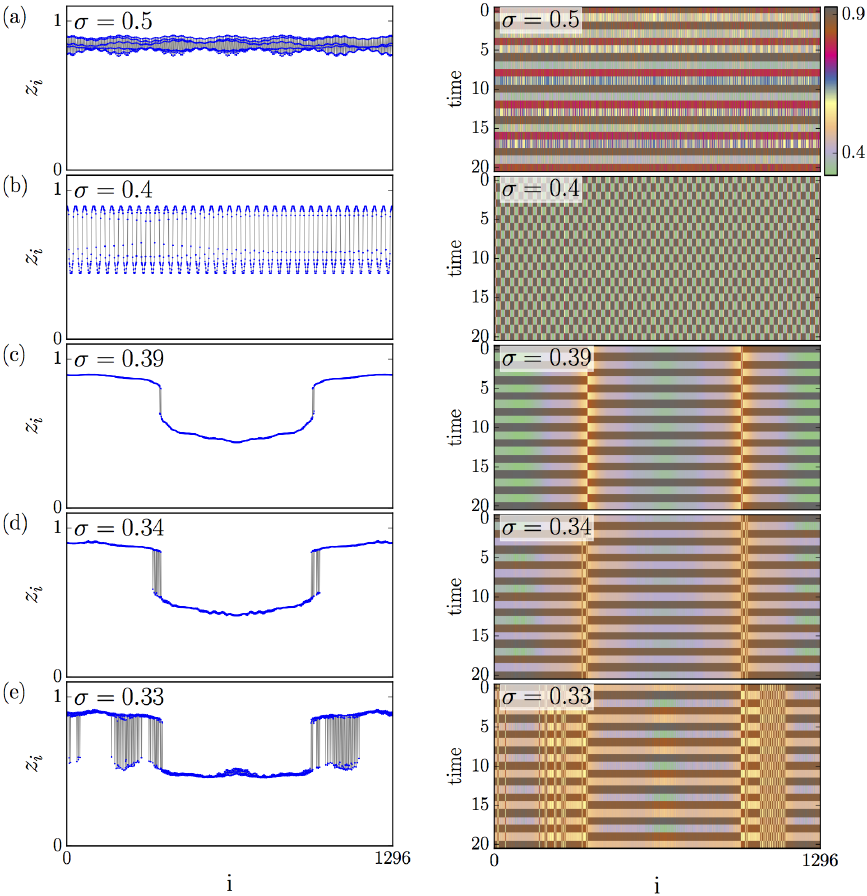}
\caption{Same as Fig.~~\ref{fig:110011m14-I} for $m=4$.}
\label{fig:110011m14-II}
\end{minipage}
\end{figure*}

We now take a closer look at the bifurcation scenarios in dependence on $\sigma$ of the networks with base patterns 110011 and 011100. For the symmetric network (110011) we compare the case with nonlocal coupling of fixed range ($m=1$) to the full hierarchical case ($m=4$) to illustrate the influence of hierarchical connectivities (Figs. \ref{fig:110011m14-I}, \ref{fig:110011m14-II}). For $m=1$ (Fig.~\ref{fig:110011m14-I}) we recover the well-known bifurcation scenario \cite{OME11,BOG16a}. For decreasing coupling strength, chimera states (b,c) develop from a coherent wave profile (a). Further decrease of the coupling strength results in an increasing incoherent domain until complex completely
incoherent patterns arise (e). 
For $m=4$ (Fig.~\ref{fig:110011m14-II}) the same scenario occurs at lower coupling strengths (c)-(e), although the coherent parts are not as smooth as for $m=1$, which is a feature commonly observed for hierarchical connectivities in this work. More differences arise for higher coupling strengths. Instead of the single-humped wave-like profile the system develops into a profile with wavenumber $k= 36$ at $\sigma = 0.4$. This profile evolves into several incoherent states up to $\sigma = 0.5$. For larger $\sigma$ the profile becomes coherent. Thus, the fractal connectivity due to a large number of connectivity gaps results in the stabilization of coherent states with large wavenumbers.

To understand the influence of the coupling term upon the dynamics of individual nodes, we rewrite equation \eqref{eq:sys} in order to separate the input of node $i$ upon itself and the input from the other elements coupled to it \cite{BOG16a}.

\begin{alignat}{3}
  z_i^{t+1} = f(z_i^t) + \frac{\sigma}{c_{n,m}} \sum_{j=0}^{N-1} C_{ij} \! \left[f(z_j^t)-f(z_i^t) \right] \\ = \underbrace{(1-\sigma)f(z_i^t)}_{\mbox{term 1}} +  \underbrace{\frac{\sigma}{c_{n,m}} \sum_{j=0}^{N-1} C_{ij}f(z_j^t)}_{\mbox{term 2}}
\label{eq:t12}
\end{alignat} 

We denote the term which depends solely on $i$ as 'term 1' and the term including the coupled nodes 'term 2'. Term 2 is recognized as average over all nodes coupled to node $i$. Term 1 is a logistic map with effective bifurcation parameter $a_{eff} = (1-\sigma)a$. Table \ref{tab:aeff} presents values of  $\sigma$ at which $a_{eff}(\sigma)$ reaches bifurcations. Within the relevant parameter range of fig.~\ref{fig:110011m14-II}, the logistic map in term 1 is operating in the nonzero fixed point or period-1 regime. For $\sigma > 0.211$ its fixed point is $z^* = 1-1/a_{eff}$ and for $\sigma > 0.737$ it is attracted to 0. Thus chaoticity or periodic dynamics with higher periods are caused by term 2 and its interplay with term 1. Term 2 is thus of interest in the analysis of the network dynamics, since its periodicity and amplitude shape the dynamics. It prevents each network element from reaching its fixed point, period-1, or period-2 cycle (if $\sigma \geq 0.1$ as usually in this work). Term 2 is an important key for the explanation of the bifurcation scenarios in dependence upon $\sigma$, at least if the system is not close to a value of $\sigma$ where term 1 exhibits a bifurcation in the parameter $a_{eff}$ (at $\sigma = 0.211$ or $\sigma = 0.737$). These bifurcations can sometimes be clearly recognized, e.g., in the dynamics of the systems with base pattern 111011. 

\begin{table}[ht]
\centering
\caption{Characteristic values of $a_{eff} = (1-\sigma)a$ as function of $\sigma$; $a = 3.8$.}
\begin{tabular}{l c r}
  \hline\noalign{\smallskip}
  $\sigma$ & $a_{eff}$ & bifurcation to...\\
\hline
0.737 & 1 & fixed point $\neq 0$\\
0.211 & 3 & period-2 cycle\\
0.092 & 3.449 & period -4 cycle \\
0.061 & 3.57 & chaos\\
  \noalign{\smallskip}\hline
\end{tabular}
\label{tab:aeff}
\end{table}

For large $\sigma \rightarrow 1$ term 1 vanishes and the dynamics of node $i$ is goverened by term 2, i.e., the average of coupled elements. For $\sigma \rightarrow 0$ the effective bifurcation parameter of term 1 $a_{eff} \rightarrow 3.8$ operates in the chaotic regime, and term 2 vanishes. If the topology is changed, only term 2 is affected.

\begin{figure}[tp]
\includegraphics[width=\columnwidth]{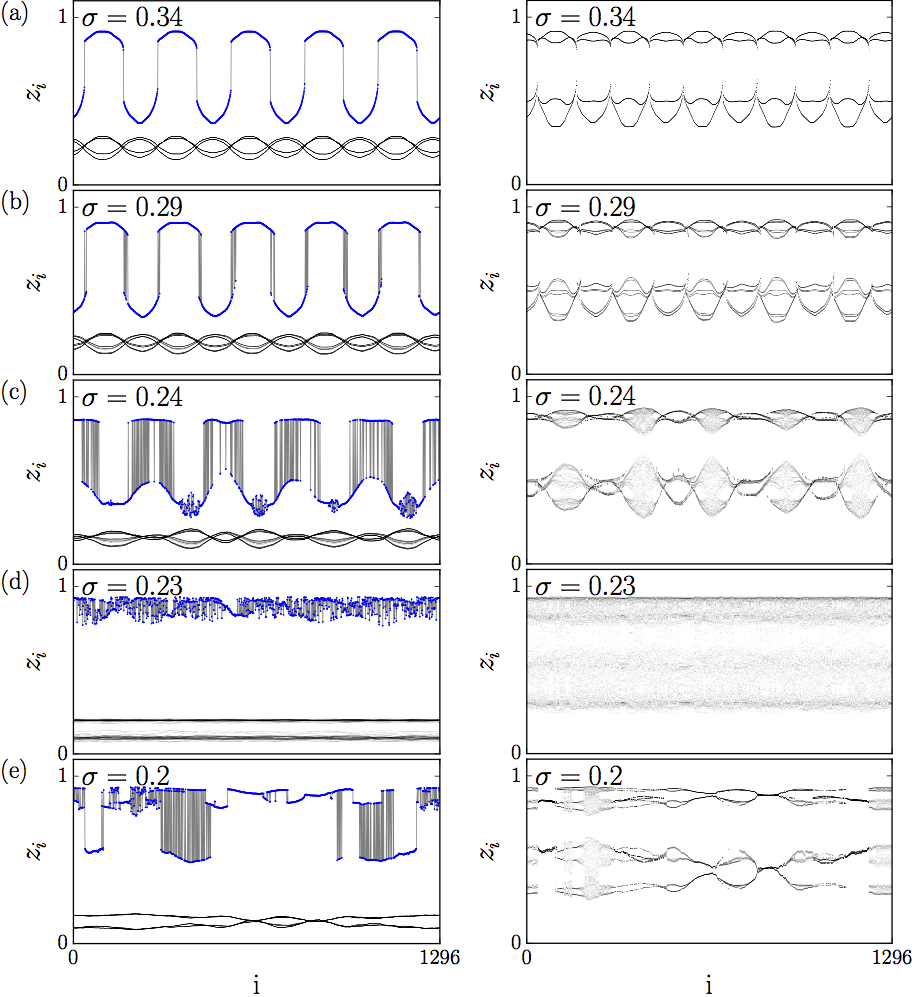}
\caption{Dynamics for base pattern 011100 and hierarchical step $m=2$. Left panel: snapshots (blue, $t=10000$) and stacked snapshots of coupling term 2 (black, last 100 iterations). Right panel: stacked snapshots of complete network (last 100 iterations). Further simulation details as in Fig. \ref{fig:110011stpprep}.}
\label{fig:c3}
\end{figure}

Figure \ref{fig:c3} shows a sequence of patterns in the network with asymmetric base pattern 011100  and hierarchical step $m=2$. To uncover the role of the coupling term, we additionally include plots of stacked snapshots of the profile, which directly show the periodicity of the dynamic. Chaotic dynamics of the coupling term for some part of the network corresponds to the appearance of small-amplitude incoherent domains. In Fig.~\ref{fig:c3} two separated chimera intervals can be recognized. With increasing $\sigma$ phase chimeras evolve into completely incoherent states (Fig.~\ref{fig:c3}(e, d)). They turn into chimera states again, now as combination of phase and amplitude chimeras (Fig.~\ref{fig:c3}(c)). With further increasing coupling strength, the incoherent domains shrink and disappear in the underlying coherent wave profile (Fig.~\ref{fig:c3}(b, a)).

Figure \ref{fig:011100stpprep} introduces a broader picture of the dynamics of systems with base pattern 011100, $m=1,2,3,4$, and a larger range of $\sigma$. For $m=3$ we encounter a similar situation as for $m=2$. A second chimera interval exists for larger coupling strengths, after the initial chimera state dissolves into an incoherent profile. This could be a signature of small multistable regions appearing close to the boundary of stability regions. For higher hierarchical step $m=4$, this second chimera interval is hardly visible.  The fragmentation of the profile for higher $m$ can be observed frequently. 
Initially coherent parts of the profile become more fragmented with increasing $m$. This is the effect of numerous connectivity gaps, and such patterns are referred to as nested chimera states~\cite{OME15}.

Compared to the symmetric base pattern 110011, in which chimeras with large incoherent domains can be observed in a large interval of coupling strength $\sigma$, base pattern 011100 displays chimeras with smaller incoherent domains over a smaller range of $\sigma$. They are more fragmented for higher hierarchical step $m$ compared to chimera observations in the symmetric pattern 110011. This can be linked to symmetry and clustering coefficient of the hierarchical connectivity. Our analysis shows that chimera states can be reliably observed in symmetric base patterns with large clustering coefficient, e.g. 110011. For higher hierarchical step $m$ the effect becomes more pronounced; albeit chimera states are found in all base patterns and for all hierarchical steps $m$ under investigation, and some asymmetric patterns display chimera states for a larger range of $\sigma$, too (e.g. 011010 and 110110).

\begin{figure*}[tb]
    \centering   
     \includegraphics[width=\textwidth]{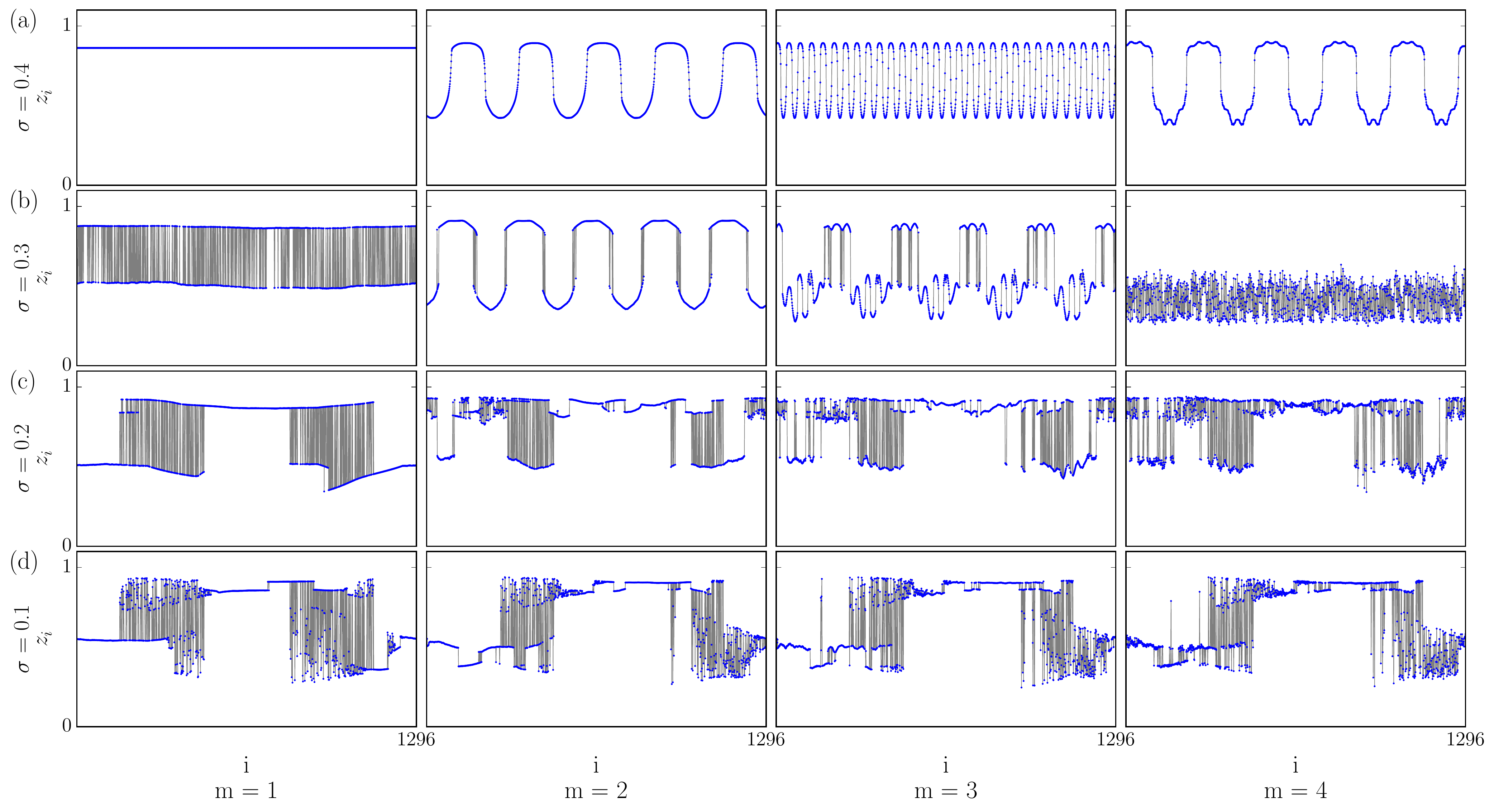}
    \caption{Snapshots of patterns for $b_{init} = 011100$, $n = 4$, at $t=10000$. Columns correspond to different hierarchical steps, coupling strength decreases from top to bottom. Further simulation details as in Fig. \ref{fig:110011stpprep}.}
    \label{fig:011100stpprep}
\end{figure*}

Finally, to uncover the role of initial conditions, we present results from numerical simulations with symmetric basestring 110011. We start from random initial conditions and $\sigma$ slightly below $\sigma_{BB}$
and successively decrease the coupling strength using the final states of the previous simulation as initial condition for the next simulation. Figs.~\ref{fig:110011rand-I} and \ref{fig:110011rand-II} present the obtained patterns for hierarchical steps $m=1$ and $m=2$. Again, we show snapshots for fixed time, stacked plots of coupling term 2 (black lines, left panels), and stacked snapshots of the whole network (right panels). In Fig.~\ref{fig:110011m14-I} we observe break-up of the smooth coherent profile with wavenumber $k=1$, and emergence of chimera states with small-amplitude dynamics. For $m=2$ a similar scenario is observed for the profile with wavenumber $k=6$. Use of random initial conditions and numerical continuation of initial conditions results in small size of the incoherent domains of phase chimeras, in correspondence with findings for networks with nonlocal coupling of fixed range~\cite{OME11,HAG12}. 



\begin{figure*}[tp]
\begin{minipage}[t]{0.48\textwidth}
\includegraphics[width=\textwidth]{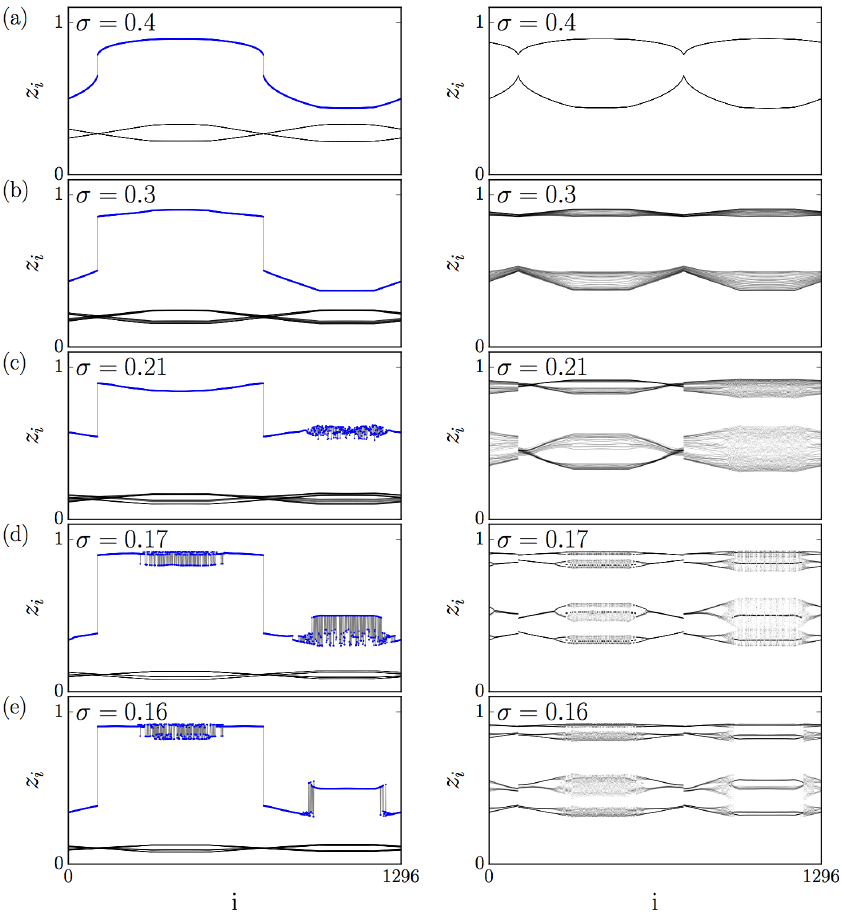}
\caption{Patterns obtained for base pattern 110011, and hierarchical step $m=1$. Left column: snapshots (blue, t=10000) and stacked snapshots of coupling term 2 (black, last 100 iterations). Right column: Stacked snapshots of complete network (last 100 iterations).}
\label{fig:110011rand-I}
\end{minipage}
\begin{minipage}[t]{0.04\textwidth}
\end{minipage}
\begin{minipage}[t]{0.48\textwidth}
\includegraphics[width=\textwidth]{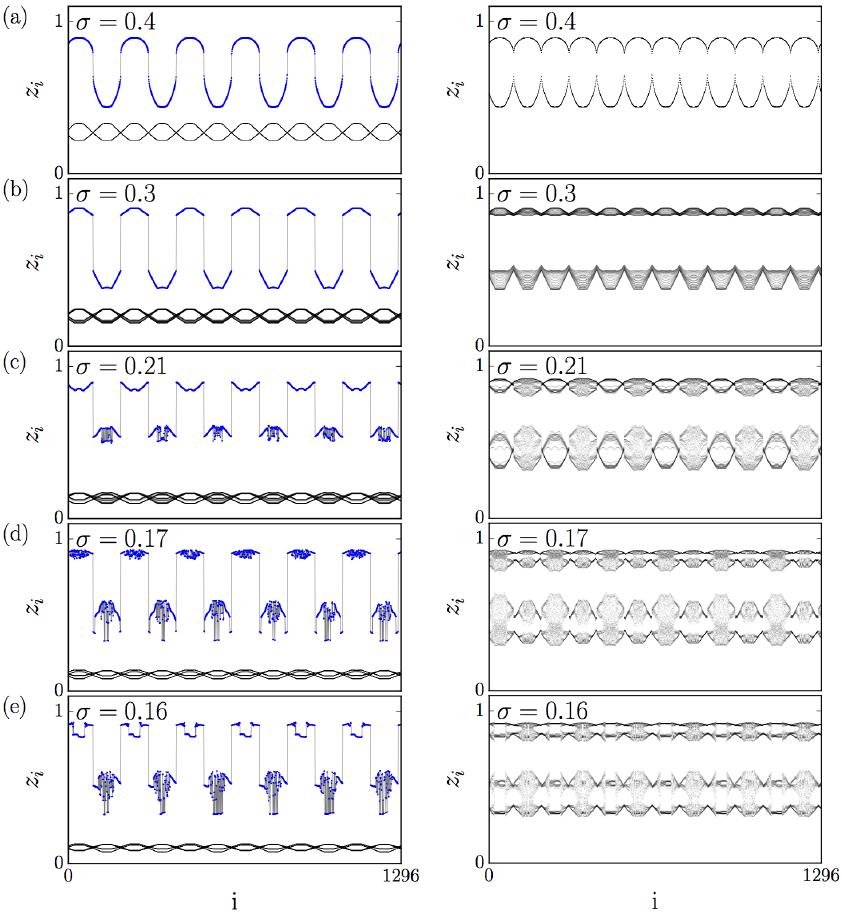}
\caption{Same as Fig.~~\ref{fig:110011rand-I} for $m=2$.}
\label{fig:110011rand-II}
\end{minipage}
\end{figure*}

Which role does term 2 play in these transitions? In our simulations we frequently observe that phase chimeras correspond to incoherent domains localized at the intersections of subsequent iterations of the map generated by term 2, Figs.~\ref{fig:c3}(b),(c) provide an example.
At these intersections the amplitude of term 2 is smaller than in the rest of the profile, and its periodicity may vary. This might be connected to the occurrence of phase chimeras because stable phase relations are disturbed. Especially if term 2 runs through a cycle of period-doubling bifurcations that causes an increasing number of intersections. 

Another hypothesis involving term 2 is that its amplitude is connected to the occurrence of small-amplitude chimeras. Figures \ref{fig:c3}, \ref{fig:110011rand-I} and \ref{fig:110011rand-II} show various examples of small-amplitude chimeras. The amplitudes are more likely to become unstable if the differences in the amplitude of term 2 are large compared to its overall level. Neighbouring nodes facing this relatively large perturbations via term 2 might be destabilized more easily, giving rise to amplitude chimeras.

\section{Conclusion}
Chimera states, i.e., patterns of coexisting coherent and incoherent domains, can be observed in networks of coupled time-discrete maps. In this case the incoherent domains of phase chimeras arise due to the random distribution of neighboring nodes between two branches of the iterated map, and many different distributions are possible due to the choice of initial conditions.

In nonlocally coupled networks of logistic maps with fixed coupling range, chimera states appear for decreasing coupling strength as a result of break-up of smooth coherent profiles. We have examined the existence and properties of chimera states in networks with hierarchical (fractal) connectivities, which are constructed by a Cantor algorithm starting from a given base pattern.  These topologies are characterized by a hierarchical structure of connectivity gaps, including long-range connections.

We have systematically analyzed possible topologies constructed from various base patterns of length $b=6$. Our analysis has shown that symmetric coupling topologies with high clustering coefficients are preferable for the occurrence of chimeras. 
We have demonstrated that chimera states are possible even for topologies with high hierarchical steps, but their diluted link structure usually results in chimeras with higher number of coherent and incoherent domains and in complex nested chimera patterns. 
For weak coupling strength we also observe small-amplitude incoherent domains associated with amplitude chimeras.
We have elaborated the influence of the coupling term on the appearance of these states.

In particular, we have compared two exemplary sets of networks constructed from symmetric and asymmetric base patterns, and have demonstrated that for the observation of regular chimera patterns symmetry and specially prepared initial conditions are crucial. Asymmetry and weak coupling results in more complex nested and small-amplitude chimera states.

Our findings may be useful for a deeper understanding of the mechanisms of formation of chimera patterns in networks with complex topologies.

\section*{Acknowledgment}
This work was supported by Deutsche Forschungsgemeinschaft in the framework of Collaborative Research Center SFB 910. 

\section*{Authors contributions}
AzB did the numerical simulations and the theoretical analysis. IO, AZ, and ES designed and supervised the study. All authors contributed to the preparation of the manuscript. All the authors have read and approved the final manuscript.

%
\bibliographystyle{ieeetr}

%
%
%

\end{document}